\documentclass[twocolumn]{aastex61}
\makeatletter

\newcommand{\Rmnum}[1]{\expandafter\@slowromancap\romannumeral #1@}
\newcommand\ha{\rm H\alpha}

\submitjournal{ApJS}
\makeatother
\usepackage{subfigure}
\usepackage{color}
\usepackage{float}
\begin{document}

\title{An $\ha$ Imaging Survey of the Low Surface Brightness Galaxies Selected from the Spring Sky Region of the 40$\%$ ALFALFA H\Rmnum{1} Survey}

\correspondingauthor{ Hong Wu}
\email{hwu@bao.ac.cn}
\shortauthors{Lei et al.}

\author{Feng$-$Jie Lei}
\affiliation{Key Laboratory of Optical Astronomy, National Astronomical Observatories, Chinese Academy of Sciences, Beijing 100012, P.R. China}
\affiliation{Nationl Astronomical Observatories, Chinese Academy of Sciences, 20A Datun Road, Chaoyang District, Beijing, 100012, China}
\affiliation{School of Astronomy and Space Science University of Chinese Academy of Sciences, Beijing 100049, China}

\author{Hong Wu}
\affiliation{Key Laboratory of Optical Astronomy, National Astronomical Observatories, Chinese Academy of Sciences, Beijing 100012, P.R. China}
\affiliation{Nationl Astronomical Observatories, Chinese Academy of Sciences, 20A Datun Road, Chaoyang District, Beijing, 100012, China}
\affiliation{School of Astronomy and Space Science University of Chinese Academy of Sciences, Beijing 100049, China}

\author{Yi$-$Nan Zhu}
\affiliation{ Schools of Physics and Astronomy, Sun Yat-sen University, Zhuhai 519082, China}

\author{Wei Du}
\affiliation{Key Laboratory of Optical Astronomy, National Astronomical Observatories, Chinese Academy of Sciences, Beijing 100012, P.R. China}
\affiliation{Nationl Astronomical Observatories, Chinese Academy of Sciences, 20A Datun Road, Chaoyang District, Beijing, 100012, China}

\author{Min He}
\affiliation{Key Laboratory of Optical Astronomy, National Astronomical Observatories, Chinese Academy of Sciences, Beijing 100012, P.R. China}
\affiliation{Nationl Astronomical Observatories, Chinese Academy of Sciences, 20A Datun Road, Chaoyang District, Beijing, 100012, China}
\affiliation{School of Astronomy and Space Science University of Chinese Academy of Sciences, Beijing 100049, China}

\author{Jun$-$Jie Jin}
\affiliation{Key Laboratory of Optical Astronomy, National Astronomical Observatories, Chinese Academy of Sciences, Beijing 100012, P.R. China}
\affiliation{Nationl Astronomical Observatories, Chinese Academy of Sciences, 20A Datun Road, Chaoyang District, Beijing, 100012, China}
\affiliation{School of Astronomy and Space Science University of Chinese Academy of Sciences, Beijing 100049, China}

\author{Pin$-$Song Zhao}
\affiliation{Key Laboratory of Optical Astronomy, National Astronomical Observatories, Chinese Academy of Sciences, Beijing 100012, P.R. China}
\affiliation{Nationl Astronomical Observatories, Chinese Academy of Sciences, 20A Datun Road, Chaoyang District, Beijing, 100012, China}
\affiliation{School of Astronomy and Space Science University of Chinese Academy of Sciences, Beijing 100049, China}

\author{Bing$-$Qing Zhang}
\affiliation{Key Laboratory of Optical Astronomy, National Astronomical Observatories, Chinese Academy of Sciences, Beijing 100012, P.R. China}
\affiliation{Nationl Astronomical Observatories, Chinese Academy of Sciences, 20A Datun Road, Chaoyang District, Beijing, 100012, China}
\affiliation{School of Astronomy and Space Science University of Chinese Academy of Sciences, Beijing 100049, China}

\begin{abstract}
We present a narrow $\ha$-band imaging survey of 357 low surface brightness galaxies (LSBGs) that are selected from the spring sky region of the 40$\%$ Arecibo Legacy Fast Arecibo L-band Feed Array (ALFALFA) H\Rmnum{1} Survey.
All the $\ha$ images are obtained from the 2.16 m telescope, operated by Xinglong Observatory of the National Astronomical Observatories, Chinese Academy of Sciences.
We provide the $\ha$ fluxes and derive the global star formation rates (SFRs) of LSBGs after the Galactic extinction, internal extinction, and [NII] contamination correction.
Comparing to normal star-forming galaxies, LSBGs have a similar distribution in the H\Rmnum{1} surface density ($\rm \Sigma_{HI}$), but their SFRs and star formation surface density ($\rm \Sigma_{SFR}$) are much lower.
Our results show that the gas-rich LSBGs selected from the ALFALFA survey obviously deviate from the Kennicutt-Schmidt law, in the relation between the star formation surface density ($\rm \Sigma_{SFR}$) and the gas surface density ($\rm \Sigma_{gas}$).
However, they follow the extended Schmidt law well when taking the stellar mass of the galaxy into consideration.
\end{abstract}

\section{Introduction} \label{sec:intro}
The low surface brightness galaxies (LSBGs) are the galaxies whose central surface brightness is at least one magnitude fainter than that of the dark sky background  \citep{1970ApJ...160..811F,1997ARA&A..35..267I}.
Because LSBGs are so faint, they are beyond the detection limit of most of the wide field optical survey \citep{1997ARA&A..35..267I}.
However, they possibly contribute 20$\%$ to the total dynamical mass of the galaxies in the universe\citep{2004MNRAS.355.1303M} and 30$\%$-60$\%$ to the number density of local galaxies \citep{1996MNRAS.280..337M, 1997PASP..109..745B, 2000ApJ...529..811O, 2006A&A...458..341T, 2007A&A...471..787H}.
To better understand LSBGs, we need to construct an appropriate sample.
\citet{2004AJ....127..704K} and \citet{2008MNRAS.391..986Z} established large LSBG samples based on the  main galaxy sample of the Sloan Digital Sky Survey (SDSS).
\citet{2015AJ....149..199D} selected 1129 H\Rmnum{1} gas-rich LSBGs by cross-matching from the SDSS data release 7 (DR7) and the Arecibo Legacy Fast Arecibo L-band Feed Array (ALFALFA) survey \citep{2005AJ....130.2598G,2005AJ....130.2613G}.

How the gas converting into the stars in galaxies is a fundamental question in galaxy formation and evolution, especially in the extremely low-density environment, such as LSBGs.
Generally, H\Rmnum{1} gas transforms into molecular gas, then collapses, and finally forms a star. 
Understanding the relationship between the star formation rate (SFR) and the gas is critical to understand the evolution of galaxies.
\citet{1959ApJ...129..243S} first proposed a relation between the SFR volume density and gas volume density.
After that, \citet{1998ApJ...498..541K} gave an empirical relation between the gas surface density ($\rm \Sigma_{gas}$) and the star formation surface density ($\rm \Sigma_{SFR}$), known as the Kennicutt-Schmidt (K-S) law:
\begin{equation}
\rm \Sigma_{SFR}\varpropto\Sigma_{gas}^{1.4}
\end{equation}

However, such an empirical relation is not suitable for dwarf galaxies or LSBGs \citep{2012AJ....143..133H,2018ApJS..235...18L}.
The gas surface density may not be the only parameter that affects the star formation.
Taking the stellar mass surface density ($\rm \Sigma_{star}$) into consideration,
\citet{2011ApJ...733...87S,2018ApJ...853..149S} proposed an extended Schmidt law:
\begin{equation}
\rm \Sigma_{SFR}\varpropto\Sigma_{gas}\Sigma_{star}^{0.5}
\end{equation}

To test the above relations in the low density environment, it is necessary to measure the correct SFR of galaxies. There are many approaches to derive the SFRs,
such as H$\alpha$, ultraviolet (UV), infrared (IR) luminosities, or fitting the observed spectral energy distribution (SED) with different models \citep{1998ARA&A..36..189K,1998ApJ...509..103S,2005ApJ...632L..79W,2008MNRAS.388.1595D,2008ApJ...686..155Z,2009A&A...507.1793N,2009ApJ...706.1527B,2014MNRAS.438...97W,2016ApJ...825...34J}.
However, the UV emission is affected by extinction, and also few LSBGs have been observed in the UV band.
The IR flux is from the dust re-emission of the light of the young massive stars. 
Unfortunately, it is also not suitable for LSBGs, because of the low dust mass of LSBGs \citep{2001ASPC..230..381M}.
The SED is a better way to get the SFR, but collecting the multi-band data simultaneously is quite challenging.

Among all the SFR tracers above, $\ha$ is a better one.
The $\ha$ luminosity is proportional to the number of newly formed stars.
It traces the stars formed over past 3-10 Myr \citep{2012ARA&A..50..531K}.
The SFR is proportional to its luminosity when the star formation activity of the target is constant on a timescale \citep{1998ApJ...498..541K}.
Generally, the $\ha$ emission of galaxies can be obtained by spectral observations (e.g., the spectroscopic survey of SDSS DR7  \citep{2002AJ....124.1810S,2003ApJ...599..971H}, and the narrow $\ha$-band imaging.
Compared to the spectroscopic observation, the narrow band imaging can obtain the total $\ha$ emission of the galaxy.

Recent $\ha$ imaging surveys of LSBGs provide resources to study their star formation.
\citet{2011AdAst2011E..12S} presented the $\ha$ imaging of 59 LSBGs selected from the Second Palomar Sky Survey (PSS-II) catalog. The $\ha3$ survey is an $\ha$ image survey of $\sim$ 800 galaxies in the Local Supercluster \citep{2012A&A...545A..16G,2013A&A...553A..89G,2015A&A...576A..16G}, which also contains some LSBGs. \citet{2018ApJS..235...18L} presented an $\ha$ survey of 111 LSBGs that are selected from the fall sky region of the $\alpha.40$ catalog, which is an H\Rmnum{1} catalog from the 40$\%$ of the ALFALFA survey area, $\sim$2800 $\rm deg^{2}$ \citep{2011AJ....142..170H}. The corresponding $\ha$ survey of LSBGs in the spring sky region still need to be completed.

In this paper, we continue to undertake $\ha$ imaging surveying of LSBGs in spring sky region of the $\alpha.40$ catalog  to explore their SFR and star formation efficiency (SFE). 
The layout of this article is as follows: in Section 2, we introduce our sample together with a description of the observations.
In Section 3 and 4, we present the data reduction and the $\ha$ flux correction.
In Section 5, we present the catalog of the $\ha$ flux and some derived parameters.
Results and a discussion are given in Section 6, and a summary is shown in Section 7.
Throughout the paper, we adopt a flat $\Lambda$CDM cosmology, with $\rm H_{0}$ = 70 km\ $\rm s^{-1}Mpc^{-1}$ and $\rm \Omega_{\Lambda}$ = 0.7, 
and a Salpeter initial mass function (IMF) $ [dN(m)/dm=-2.35]$ over $ m=0.1-100 M_{\sun}$ \citep{1955ApJ...121..161S}.


\section{Sample and Observations}

\begin{figure}[!ht]
\includegraphics[scale=0.3]{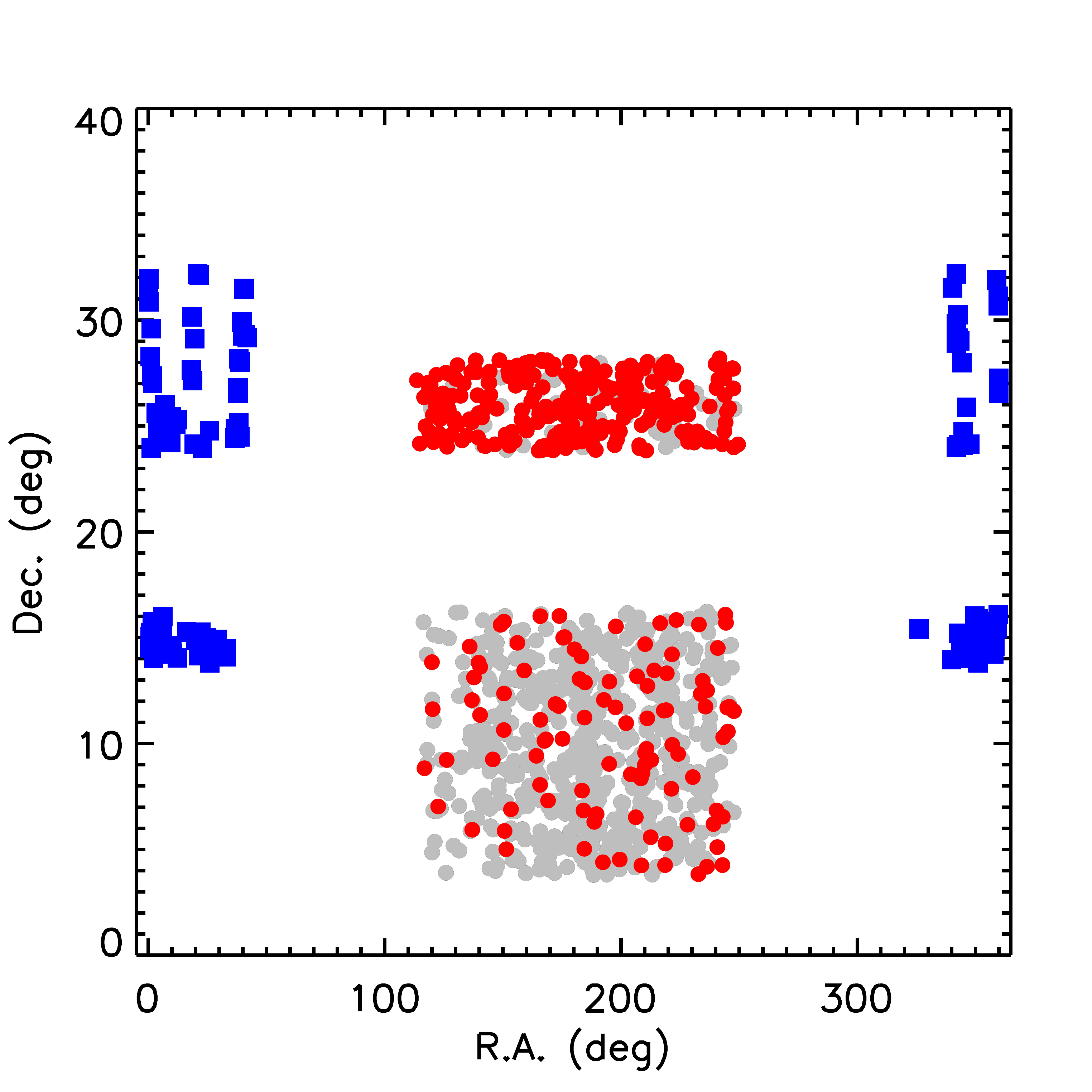}
\caption{Sky distribution of the LSBGs. 
The solid circles and squares are the 1129 LSBGs from \citet{2015AJ....149..199D}.
The blue squares refer to  the observed LSBGs in the fall sky region.
The red solid circles are  the observed LSBGs in the spring sky region.
The others (gray solid circles) are the unobserved objects due to the limitation of the observation time.
}
\label{radec}
\end{figure}

\begin{figure*}[!ht]
\centering
\includegraphics[scale=0.3]{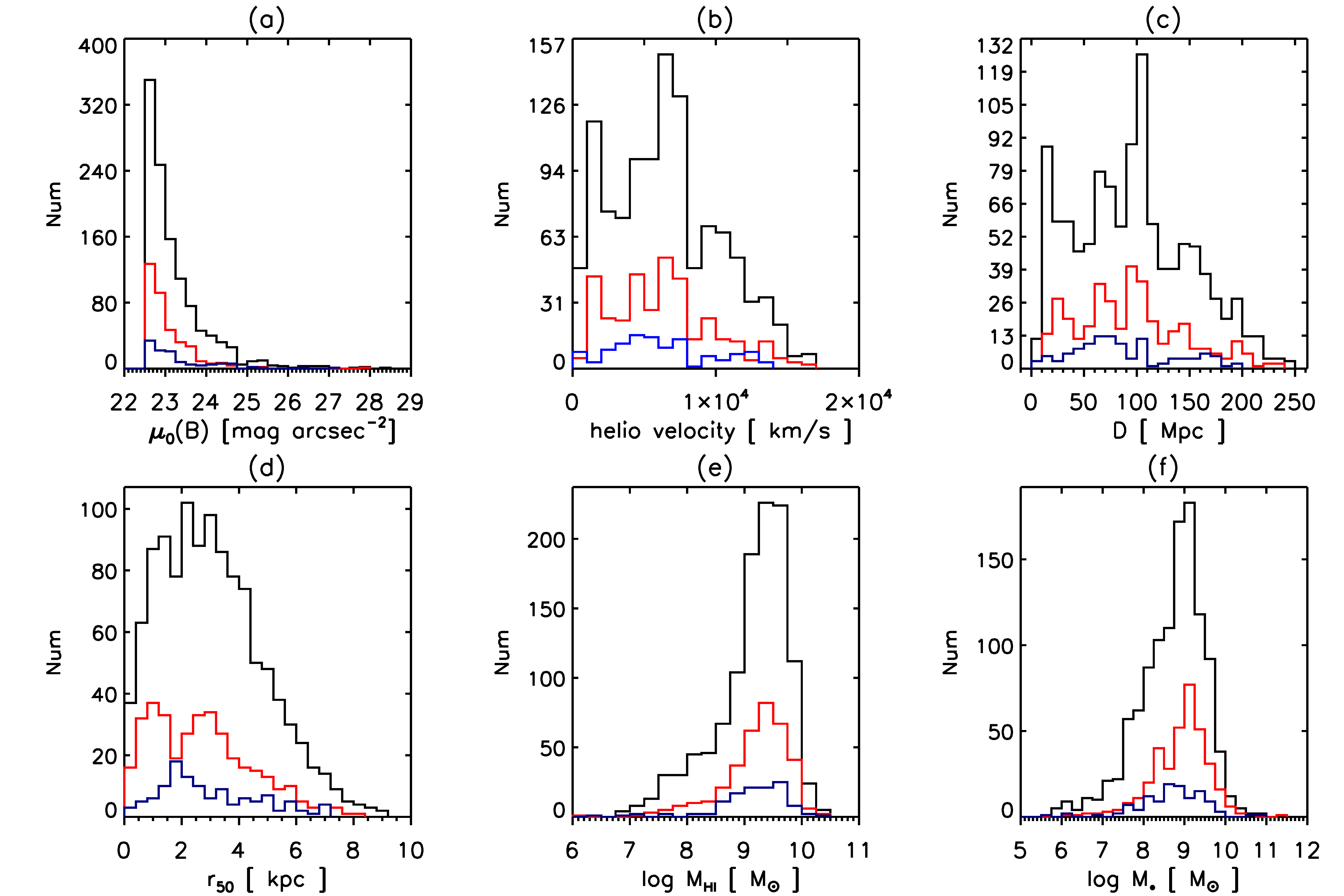}
\caption{Histograms of six parameters of the observed LSBGs in the fall (blue) and spring (red) sky region and the whole LSBGs of Du2015 (black).
(a): Central surface brightness in the B band with a bin size of 0.25 .
(b): Heliovelocity of an H\Rmnum{1} source in units of $\rm km~s^{-1}$ .
(c): Distance in Mpc from the $\alpha$.40 catalog \citep{2011AJ....142..170H}. 
(d): Radii at a 50$\%$ fraction of light in r band in units of kpc.
(e): H\Rmnum{1} mass from $\alpha$.40 catalog \citep{2011AJ....142..170H}.
(f): Stellar mass derived from g-r color and r-band luminosity \citep{2003ApJS..149..289B}.
}
\label{dispar}
\end{figure*}

\citet{2015AJ....149..199D} selected 1129 LSBGs from the $\rm \alpha$.40-SDSS-DR7 with the B-band central surface brightness $\rm \mu_{0}(B)$ fainter than 22.5 $\rm mag\ arcsec^{-2}$ and the axis ratio of b/a $>$ 0.3.
$\alpha$.40 \citep{2011AJ....142..170H} is the first released H\Rmnum{1} catalog of the Arecibo Legacy Fast ALFA(ALFALFA) survey \citep{2005AJ....130.2598G,2005AJ....130.2613G} and covers a 40$\%$ area of a total of 7000 $\rm deg^{2}$.
The SDSS DR7 images, whose sky background are overestimated by 0.2-0.5 mag \citep{2007ApJ...660.1186L,2013ApJ...773...37H}, bring the challenge to search for LSBGs.
\citet{2015AJ....149..199D} rebuilt the sky background of SDSS images, and fitted the galaxies with the exponential disk model using the GALFIT software, and derived more accurate $\rm \mu_{0}(B)$.
They constructed a sample of 1129 LSBGs, hereafter Du2015.
Since Du2015 selected LSBGs sample from the ALFALFA H\Rmnum{1} survey, this could make the sample slightly H\Rmnum{1}-rich biased.

The sky distribution of the LSBGs from Du2015 is shown in Figure \ref{radec}.
Based on Du2015, we observed 468 $\ha$ images of LSBGs.
\citet{2018ApJS..235...18L} presented the first results of 111 LSBGs in the fall sky region  of Du2015,  which are shown as the blue squares. 
Due to the limited observation time, the observation of the spring sky region of Du2015 is not finished.
The spring LSBGs sample covers two sky fields.
We first focus on the smaller upper field at the decl. of $\sim 25^{\degr}$, so most of the LSBGs in this field are observed.
 The LSBGs in the larger field at the decl. of $\sim 10^{\degr}$ are observed randomly. 
Therefore, our observation does not introduce any selection effect to the spring LSBG sample. 
A total of 357 spring LSBGs of Du2015 observed are shown as the red solid circles in the figure.   

As a comparision, we present the distributions of some fundamental  parameters of the observed LSBGs in the fall sky fields (blue), the spring sky fields (red), and all the LSBGs(black) of Du2015 in Figure \ref{dispar}. In general, the distributions of the six parameters of the observed spring LSBGs agree well with those of Du2015.
So we believe the observed spring LSBG sample can represent the whole LSBG sample of Du2015 without bringing in the selection bias from the observation.

Our observation includes not only the narrow $\ha$-band image but also the broad R-band image.
The broad R-band is used as the auxiliary image of continuum to be subtracted from the $\ha$ image.
All the $\ha$ and R images are observed using BFOSC instrument BAO Faint Object Spectrograph and Camera(BFOSC) attached to the 2.16 m telescope \citep{2016PASP..128k5005F} at the Xinglong Observatory of the National Astronomical Observatories, Chinese Academy of Sciences (NAOC).
The effective wavelength $\rm \lambda_{\mathrm{eff}}$ of the broad R-band filter is 6407$\rm \AA$ with a FWHM of 1200 $\rm \AA$. 
There are 11 $\ha$ filters whose central wavelengths  range from 6533 to 7052 $\rm \AA$ with a FWHM of $\sim$ 55 $\rm \AA$.
We only use filters of $\ha$1-7. More detailed information about filters can be found in \citet{2018ApJS..235...18L}.

The $\ha$ image observation is from 2013 to 2017.
Most of the observations are under photometric conditions. 
The exposure time are 300s for the R-band and 1800s for the narrow $\ha$ band, respectively.
To save observation time, we did not observe the standard stars.
All the observation information about 357 spring LSBGs is listed in Table \ref{tab:tableone}, including name, magnitude, coordinates, distance, filter, and observation date.

\section{image reduction}
\noindent{\emph{General Image Processing}}\\
The general image data reduction includes overscan correction, bias subtraction, image trimming, flat-field correction, cosmic-rays removal, world coordinate system (WCS) calibration and background subtraction.
The charge-coupled device (CCD) reduction (overscan, bias and flat-field) are done following the standard IRAF  procedures .
Cosmic rays are removed by using the IDL program la$\_$cosmic.pro \citep{2001PASP..113.1420V}.
A celestial coordinate is added into the image FITS header with the help of Astrometry.net.

We subtract the background as follows.
 Firstly, we produce the object-masked image by using SExtractor software to detect objects in the gauss smoothed image.
It is much easier to detect the extend wings of bright stars and the fainter outer parts of the galaxies in the gaussian-smoothed image than in the original image.
According to the detected regions in the Gauss smoothed image, we mask all the objects in the original image.
Then, a median filter of $\rm 70 \times 70\ pixel^{2}$ is applied to the object-masked image. 
 The median filter can fill in the mask regions with the surrounding sky background, and the sky background image can be obtained.
Finally, we subtract the background from original image.   
The mean value of the final sky-subtracted image is close to 0, and the fluctuation is smaller than that of the original image. An example is shown in Figure 4 of \citet{2018ApJS..235...18L}.\\
\hspace{1cm}\\
\emph{Continuum Subtraction}\\
\begin{figure}[!ht]
\centering
\includegraphics[scale=0.5]{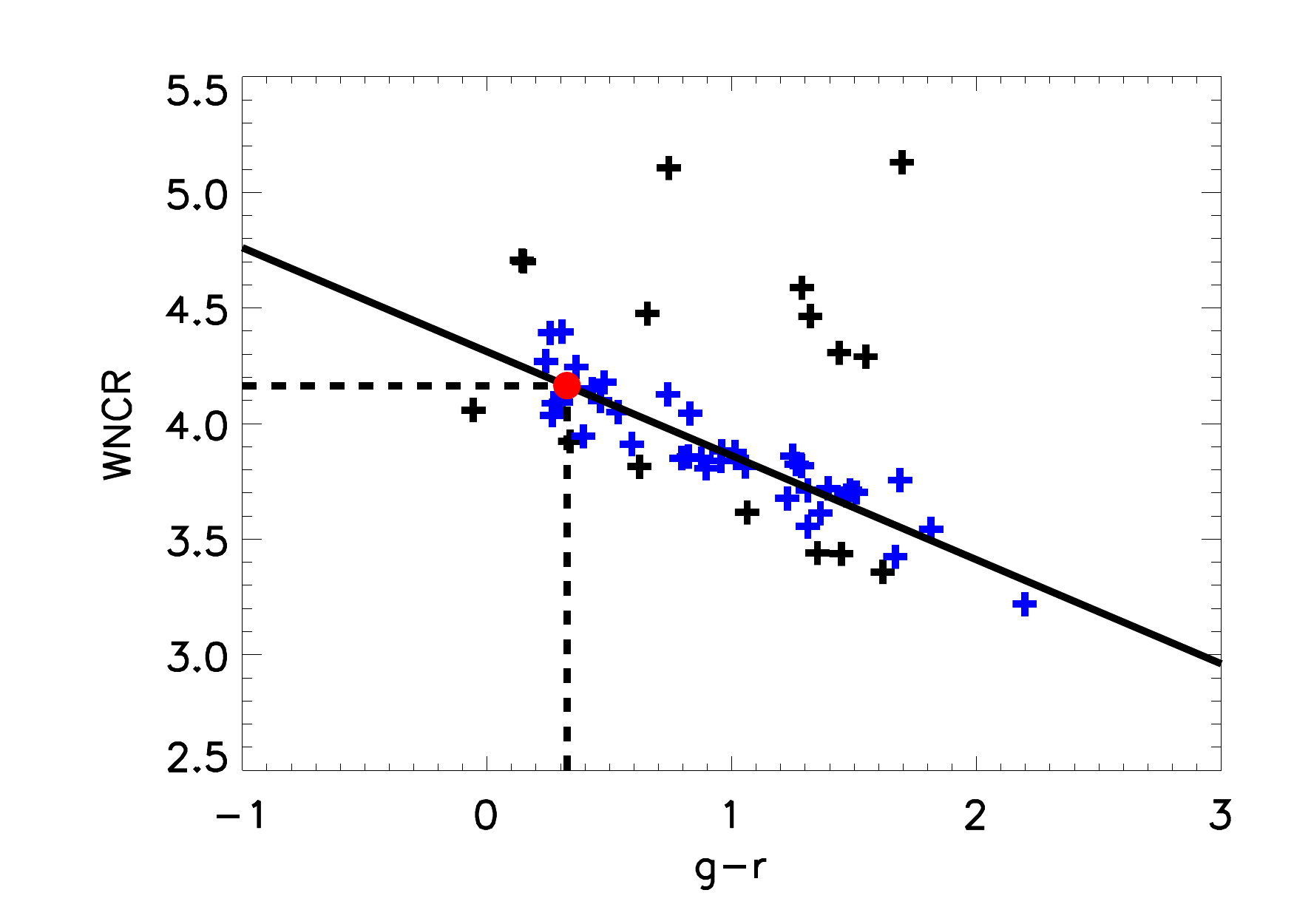}
\caption{Color effect. The calculated WNCRs of the field stars in an example image as a function of their $g-r$ colors.
The blue pluses are stars after  $1\sigma$ clipping (black pluses) and are used to fit the line.
Given the color of target galaxy (red solid circle), the WNCR can be derived from the fitting.
 }
\label{subcon}
\end{figure}
\begin{figure*}[!ht]
\centering
\includegraphics[scale=0.25]{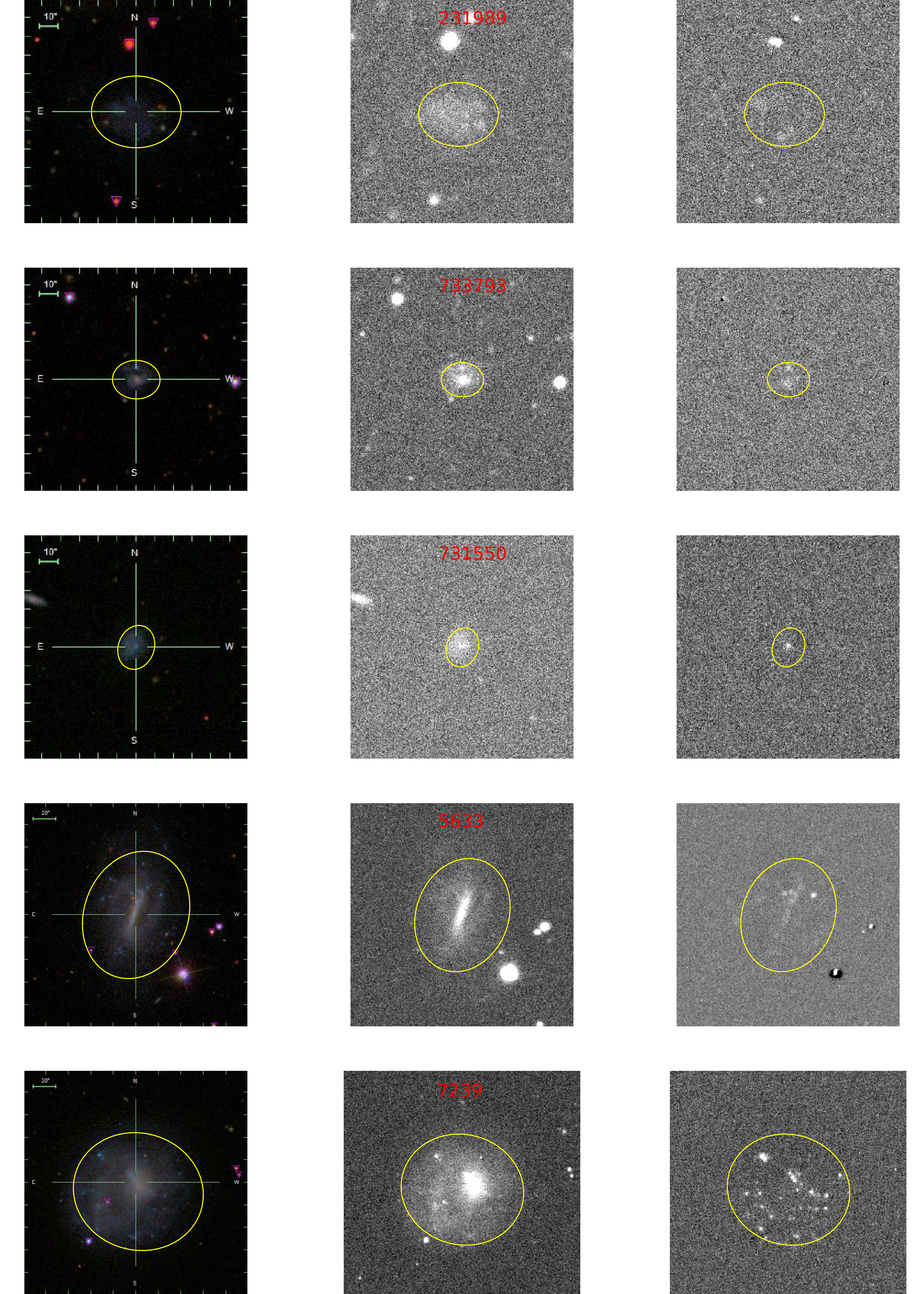}
\caption{ Here shows the SDSS rgb images, the R-band images, and the continuum-subtracted $\ha$ images of five representative galaxies from left to right.
The yellow ellipses are the photometric apertures.
}
\label{photo}
\end{figure*}
\begin{figure}[!ht]
\centering
\includegraphics[scale=0.5]{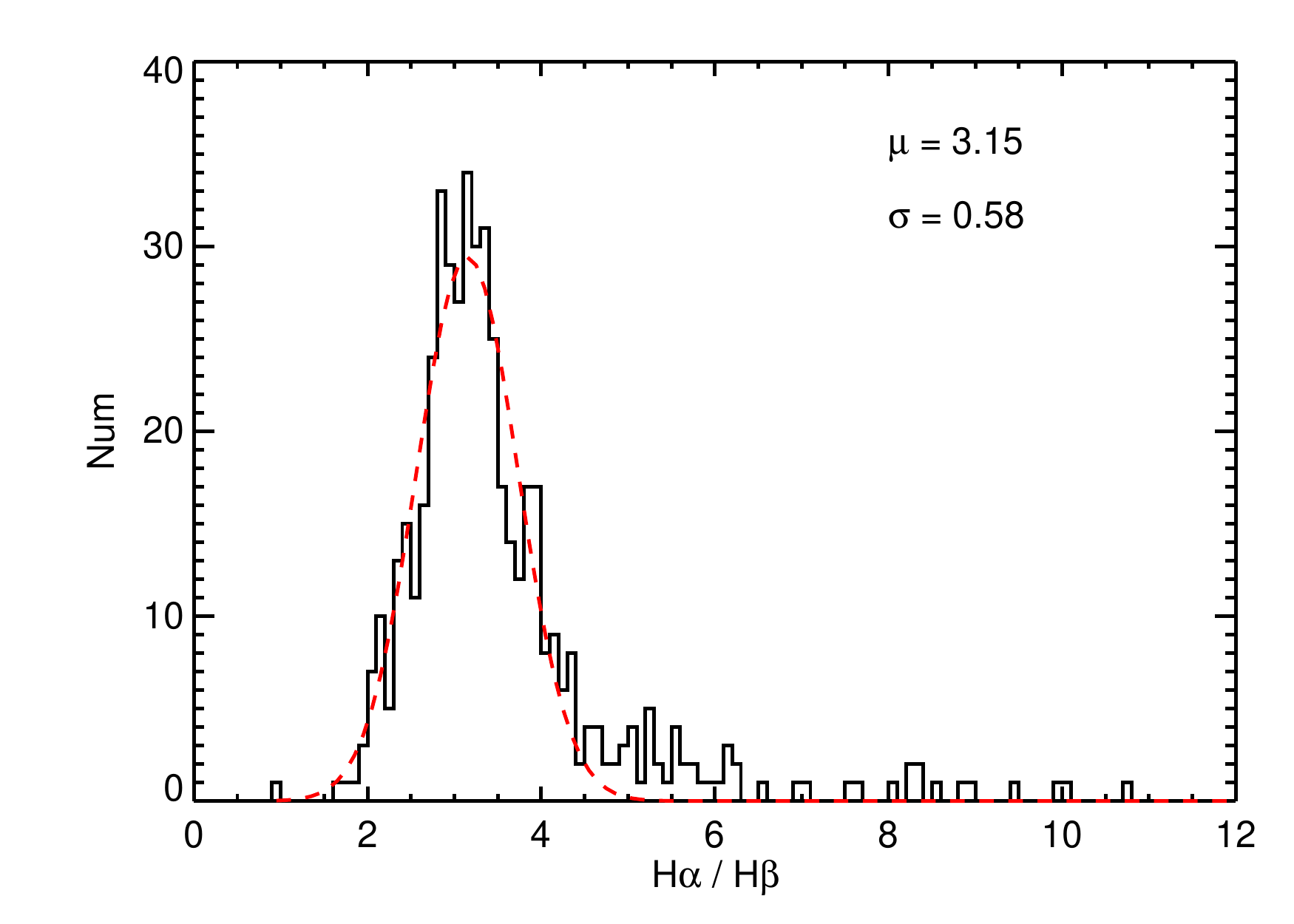}
\caption{ Distribution of $\rm F_{H\alpha}/F_{H\beta}$ of 510 LSBGs with SDSS spectra.
 }
\label{balmer}
\end{figure}
\begin{figure}[!ht]
\centering
\includegraphics[scale=0.7]{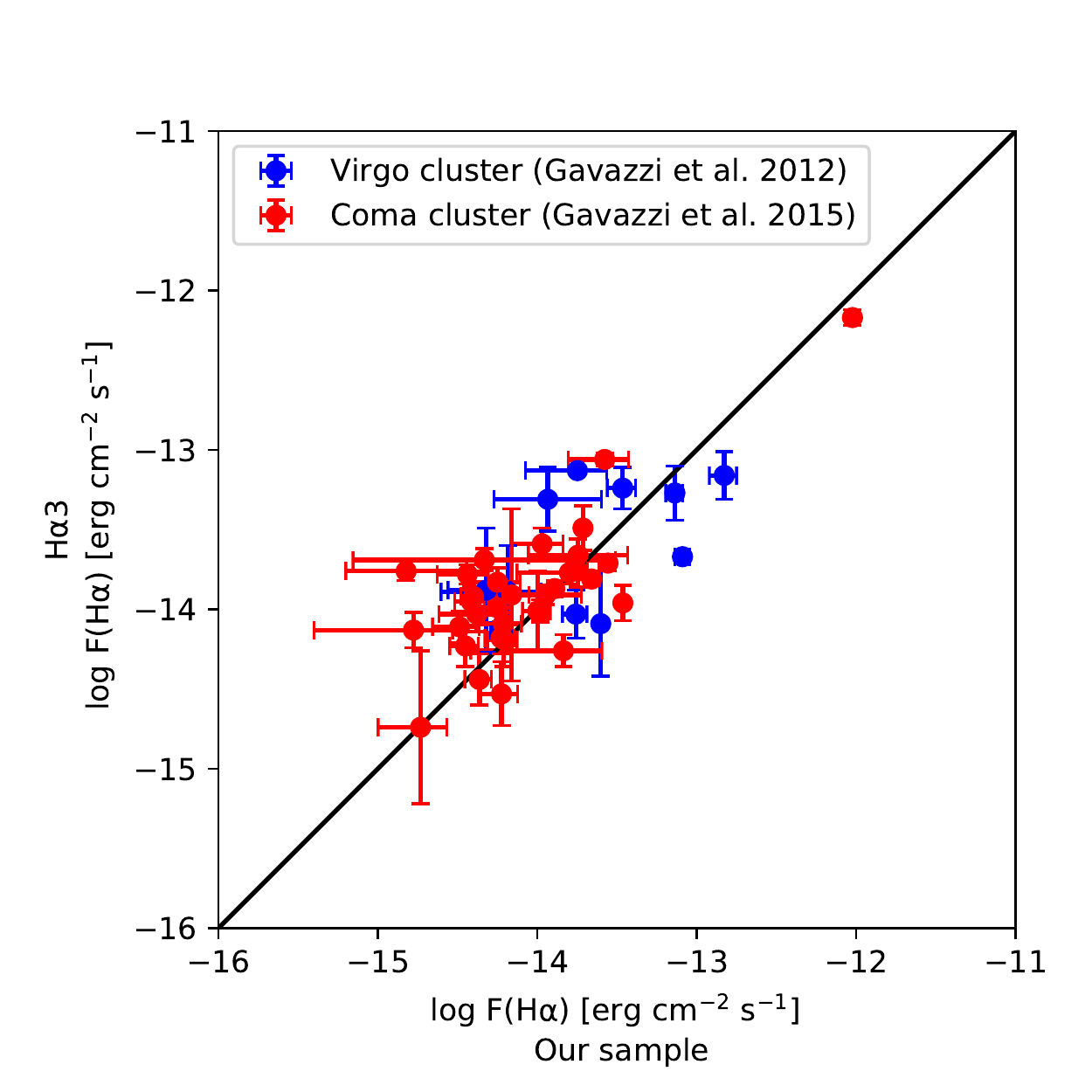}
\caption{Comparison of the $\ha$ flux of the common LSBGs from our sample and the $\ha3$ survey. The blue solid circles are galaxies matched with the Virgo cluster \citep{2012A&A...545A..16G}. The red solid circles are galaxies matched with the Coma cluster \citep{2015A&A...576A..16G}. The error bars of the $\ha$ flux are from both our and the $\ha3$ measurements.
 }
\label{comflux}
\end{figure}
\begin{figure*}[!ht]
\centering
\includegraphics[scale=1]{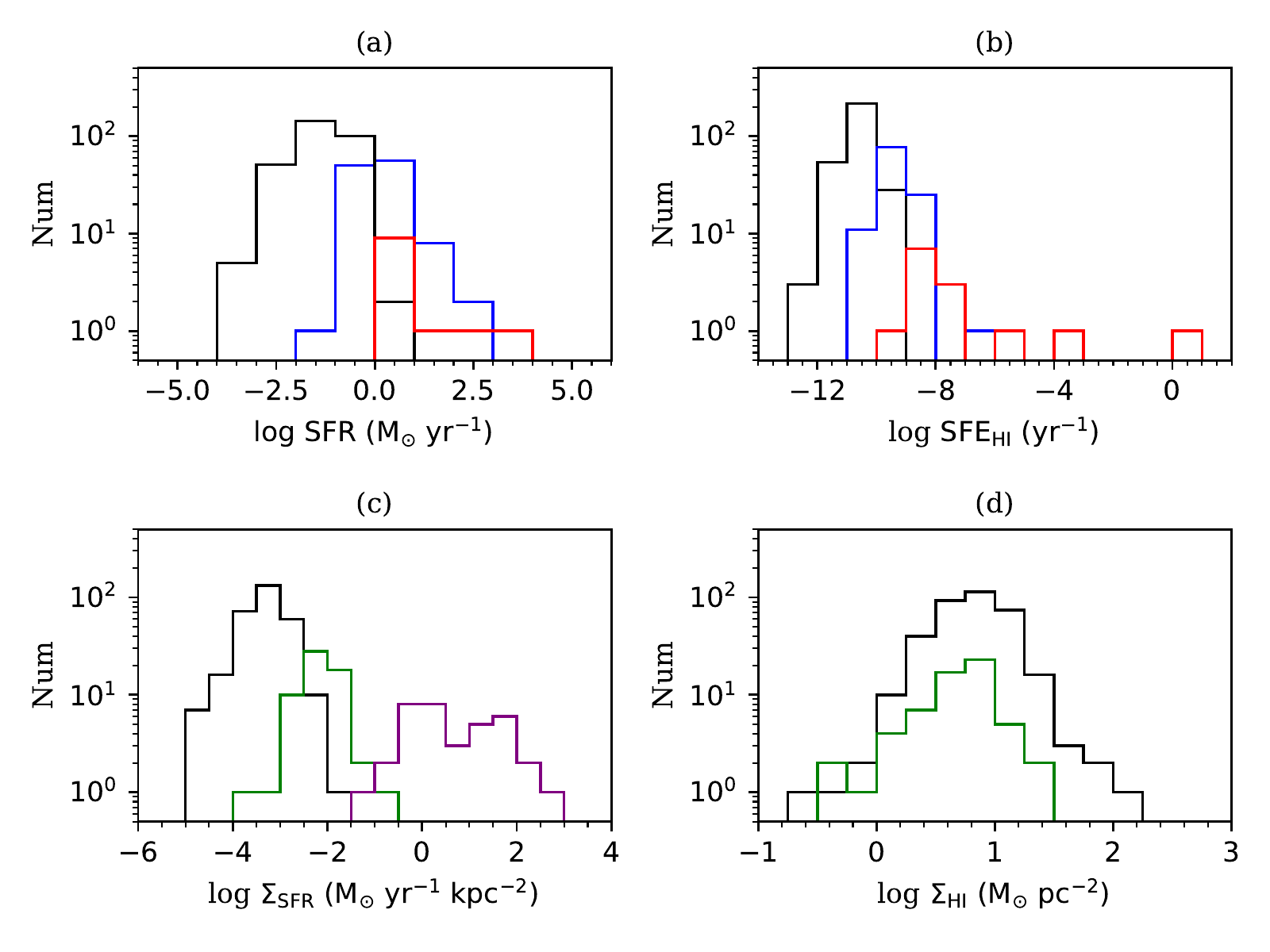}
\caption{
Distributions of the (a) SFR, (b) star formation efficiency of H\Rmnum{1}, (c) star formation surface density  and (d) H\Rmnum{1} mass surface density.
Black lines show the corresponding distributions of our observed spring LSBGs.
In panel (a) and (b), the blue and red lines show the distributions of the star-forming galaxies and starburst galaxies from \citet{1996AJ....112.1903Y} and \citet{2015ApJ...808...66J}, respectively.
In panel (c) and (d), the green and purple lines are the star-forming galaxies and starburst galaxies from \citet{1998ApJ...498..541K}.
}
\label{segsfr}
\end{figure*}
Since $\ha$ images contain the contributions from both the $\ha$ emission and the underlying stellar continuum.
The pure $\ha$ emission image can be obtained by subtracting the R-band image from the scaled  $\ha$ image. 
It is very important to determine the scale parameter.
As the field stars have no emission in the observed $\ha$ filter, their ratios of continuum fluxes in the R-band image to those in the narrow $\ha$ image can be used as the scale parameter:
\begin{equation}
WNCR = \frac{c_{W}}{c_{N}}
\end{equation}
where $c_{W}$ and $c_{N}$ are the measured fluxes of the field stars in the wide R band and narrow $\ha$ band, respectively.

There are two ways to obtain the final scale factor for a target galaxy.
One way is to adjust the wide to narrow continuum ratio (WNCR) value in a reasonable range,
and adopt the best WNCR value when the residual of fluxes of most field stars reaches a minimum \citep{2008ApJS..178..247K,2018ApJS..235...18L}.
However, this does not take the color effect \citep{2012MNRAS.419.2156S,2018A&A...611A..28G} into account. 

Because the effective wavelengths of the broad R band and the narrow $\ha$ band are different, the WNCR (the ratio of the integral continuum in the R band to those in the $\ha$ band) is related to the slope of the continuum of the target in broad R band. This is the color effect. The slope of the continuum can also be described by the color (e.g., g-r). 
The value of the WNCR may correlate with the color. 
However, the color of a galaxy is often different from the colors of most field stars. 
This could lead to the underestimation of $\ha$ flux as large as 40$\%$ and the overestimation as large as 10$\%$  \citep{2012MNRAS.419.2156S}.
Another way to get WNCR value is to take the color of the target galaxy into account. 
As an example shown in  Figure \ref{subcon}, a linear fitting is applied for the relation of WNCRs and colors of the field stars after the 1$\sigma$ clipping.
We derive the WNCR value according to the color of the target galaxy from the fitting line. 
Once the WNCR is determined, the pure $\ha$ emission image can be obtained by subtracting R-band image from the scaled $\ha$ image.\\
\hspace{1cm}\\
\emph{Flux calibration}\\
 Because all the sample LSBGs are selected from the SDSS imaging survey, we can do the flux calibration by SDSS photometry. We first extract the  r- and i- aperture magnitudes from SDSS for all the field stars in each observed field. Then the SDSS r- and i-band magnitudes are transferred  into the Johnson R-band magnitude according to  \citet{2005AAS...20713308L}.
From the aperture-measured ADUs of each field star in the R-band image and the derived Johnson R-band flux, we obtain the calibration coefficient. 
Finally, the most probable value of the coefficients of the stars in each field is adopted to calibrate the corresponding target galaxy. \\
\hspace{1cm}\\
\emph{Photometry}\\
Photometry is performed with the aid of the ellipse package of the IRAF.
It uses the elliptical isophotes to fit a galaxy and outputs the photometric and geometric parameters.
To be consistent with the radius of the surface density of the SFR in \citet{1998ApJ...498..541K}, $r_{25}$ is adopted as the photometric radius.
We use the R-band image to determine the photometric radius $r_{25}$, where the surface brightness magnitude reaches to 25 $\rm mag\ arcsec^{-2}$.
Figure \ref{photo} shows the SDSS rgb images, the R-band images, and the continuum-subtracted $\ha$ images of five representative LSBGs from left to right.
The yellow ellipses are photometric aperture.
The $\ha$ flux is the total flux within the $r_{25}$ elliptical area.
\\
\hspace{1cm}\\
\emph{Errors}\\
There are two major errors in the data reduction and photometry. 
One is the photometric error, and the other is the error from the continuum subtraction.
The photometric error consists of photonic noise of $\ha$ emission and all the statistical noise from the background, including the noise from the CCD and sky background.
The uncertainty of the WNCR in continuum subtraction is another important source of the error.
Since the WNCR is the flux ratio of the R band to the $\ha$ band, the systematic deviation of continuum subtraction could be the dominant error for the galaxies with a strong continuum and relatively weak $\ha$ emission \citep{2008ApJS..178..247K}.
The final error {is composed of} both errors of the photometry and continuum subtraction and is listed in Table \ref{tab:tabletwo}.


\section{Flux Correction}

The extinction plays an important role in determining the accuracy of the SFRs.
The extinction includes the Galactic and intrinsic extinction.
Because SDSS r-band filter covers $\ha$ emission line, we adopt the extinction value of the SDSS r-band to correct the Galactic extinction of  the observed $\ha$ emission.

Generally, the intrinsic extinction correction is derived from the Balmer emission line ratio of $\rm F_{H\alpha}/F_{H\beta}$. 
We adopt the intrinsic ratio of $\rm F_{H\alpha0}/F_{H\beta0}=2.87$.
The color excess $\rm E(B-V)$ can be derived by the Cardelli Clayton Mathis (CCM) extinction law, which is applicable to both diffuse and dense regions of the interstellar medium \citep{1989ApJ...345..245C}.
The extinction can be calculated from $\rm R_{V} = A_{V}/E(B-V)=3.1$, and $\rm A_{\ha}/E(B-V)=2.468$ \citep{2001PASP..113.1449C}.
There are 510 LSBGs of Du2015 whose  SDSS spectra data are available. The distribution of  $\rm F_{H\alpha}/F_{H\beta}$ of 510 LSBGs is shown in Figure \ref{balmer}. As the $\rm F_{H\alpha}/F_{H\beta}$ does not depend on either the central surface brightness or the color of $g-r$, 
we finally adopt the gaussian fitting value of $F_{H\alpha}/F_{H\beta}=3.1493$ (with 1$\sigma$ of 0.58) of 510 LSBGs to do the extinction correction for those  without SDSS fiber spectra. 

$\rm [NII]$($\lambda\lambda$6548, 6584) also contribute to the $\ha$ images.
We can remove these $\rm [NII]$ contributions by the ratio of $\rm [NII]$/$\ha$.
\begin{equation}
\rm f_{\ha,corr[NII]}=\frac{f_{\ha+[NII]}}{1+\frac{f_{[NII]}}{f_{\ha}}}.
\end{equation}
For the LSBGs whose SDSS fiber spectra are not available, we similarly take the median ratio of $\rm[NII]/\ha = 0.1578$ to correct the contamination from $\rm [NII]$ emission.

Taking the transmission curve of the $\ha$ filters  into account, we adopt the transmission curve of $\ha$ filters \citep{2018ApJS..235...18L} and correct the transmission loss at the wavelength of the redshifted $\ha$ line of the target galaxy.
The normalized transmission $\rm T(\ha)$  used for the flux correction is derived from the equation bellow:
\begin{equation}
\rm T(\ha) =\frac{ T'(H\alpha)}{  \int_{\lambda1}^{\lambda2}{T'(\lambda)d\lambda}/ FWHM}
\end{equation}
where $\rm T'(\lambda)$ is the transmission curve,
$\rm T'(H\alpha)$ is the direct transmission at redshifted $\ha$ line from the transmission curve,   
$\rm T(H\alpha)$ is the normalized transmission at redshifted $\ha$ ,
$\lambda1$ and $\lambda2$ are the starting and ending wavelength of the transmission curve.
The FWHM is the full width at half maximum of the $\ha$ filters.
The transmission curve $\rm T'(\lambda)$ and FWHM of each $\ha$ filter can be found in \citet{2018ApJS..235...18L}.
The corrected $\ha$ flux is obtained after divided by the normalized transmission $\rm T(\ha)$.

The R-band flux also contains the contribution from the $\ha$ emission, which will result in the loss of $\ha$ emission flux during the process of stellar continuum subtraction.
Fortunately, such a loss can be estimated (about 4$\%$) and corrected according to the FWHMs of both the R (1200 $\rm \AA$) and the $\ha$ (55 $\rm \AA$) filters.


\begin{figure*}[!ht]
\centering
\includegraphics[scale=0.65]{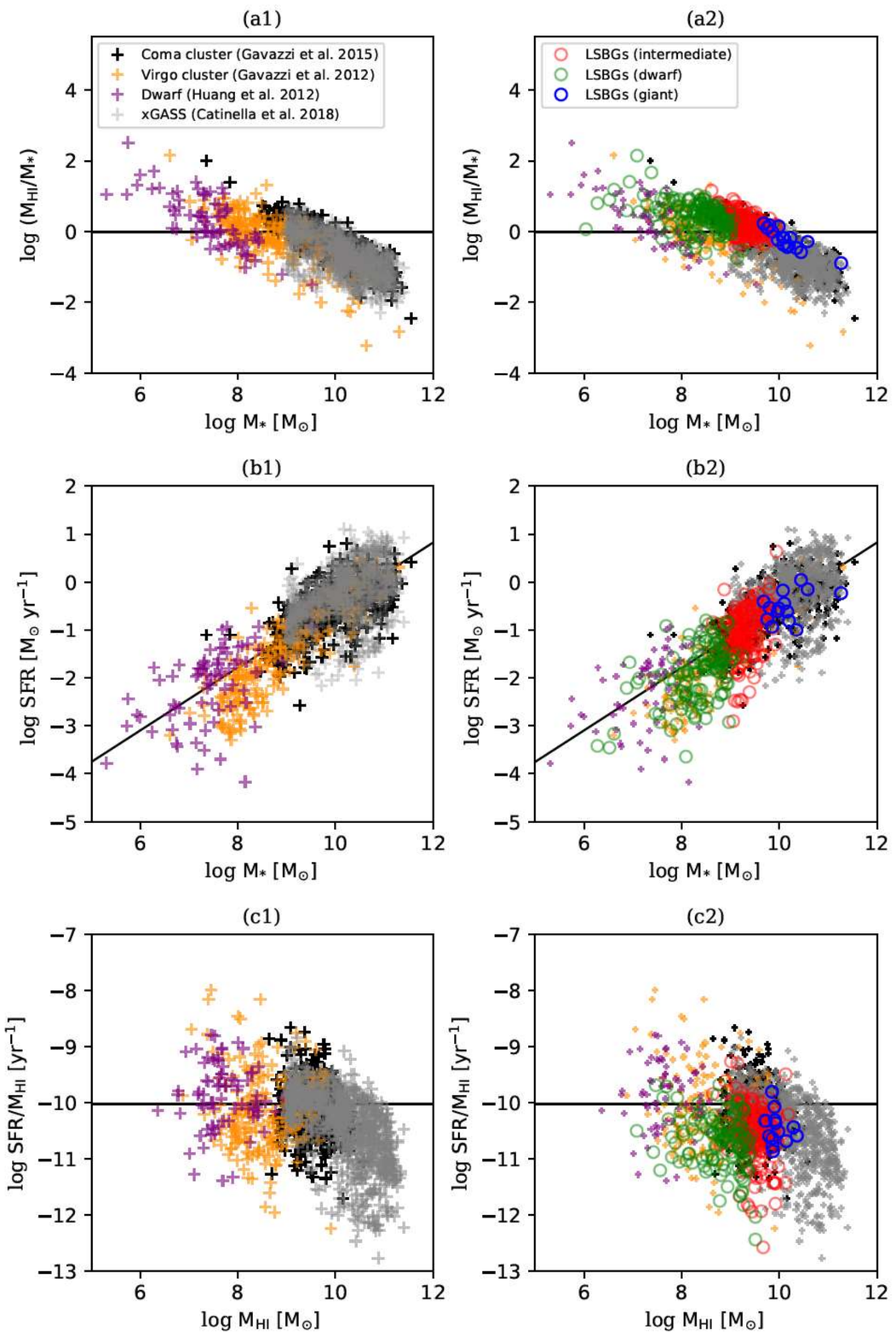}
\caption{ 
(a1), (a2): $\rm M_{HI}/M_{*}$ vs. $\rm M_{*}$. (b1),(b2): the SFR vs. $\rm M_{*}$. (c1),(c2): $\rm SFR/M_{HI}$ vs. $\rm M_{HI}$.
The left panels present the comparison galaxies, which are the dwarf galaxies (purple pluses), the galaxies in the Coma (black pluses) and Virgo clusters (brown pluses) from the $\ha3$ survey and the galaxies (gray pluses) from xGASS.
The right panels show three types of our LSBGs: the giant LSBGs (blue open circles ), the intermediate LSBGs (red open circles), and the dwarf LSBGs (green open circles)\citep{2019MNRAS.483.1754D}, and the comparison galaxies are shown in smaller pluses.
In panel (a), the solid line represents when the H\Rmnum{1} mass $\rm M_{HI}$ equals the stellar mass $\rm M_{*}$.
In panel (b), the solid line is the fitting line of the stellar mass $\rm M_{*}$ vs. SFR  derived from the comparison galaxies.
In panel (c), the solid line is the median value of log $\rm SFR/M_{HI}$ of comparison galaxies.
}
\label{sfrmass}
\end{figure*}

\begin{figure*}
\centering
\includegraphics[scale=0.7]{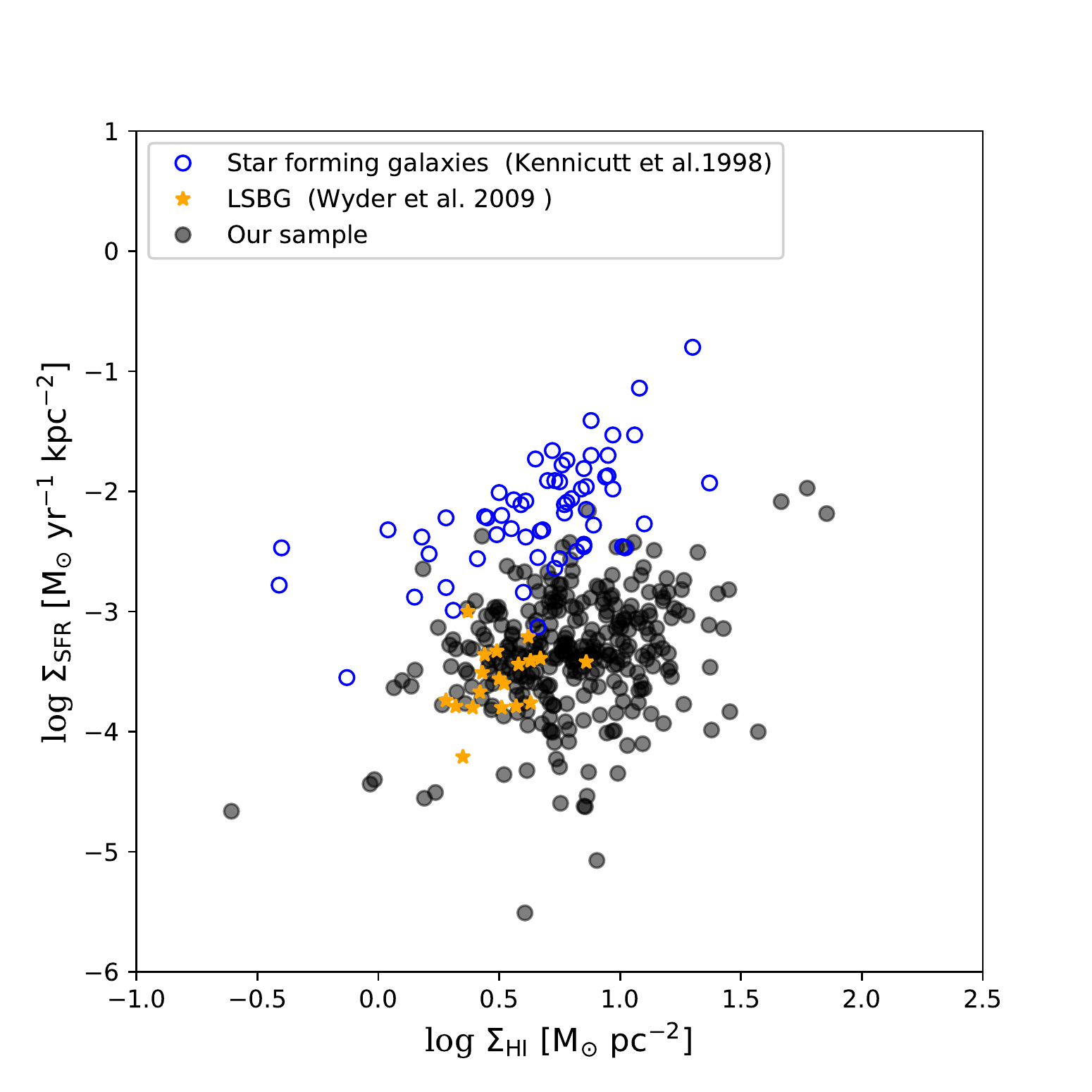}
\caption{The relation between the SFR surface density and H\Rmnum{1} gas surface density.
Our LSBGs sample are the black solid circles.
The blue open circles are the star-forming galaxies from \citet{1998ApJ...498..541K}.
The orange stars are the LSBGs from \citet{2009ApJ...696.1834W}.
}
\label{hiks}
\end{figure*}

\begin{figure*}
\centering
\includegraphics[scale=0.7]{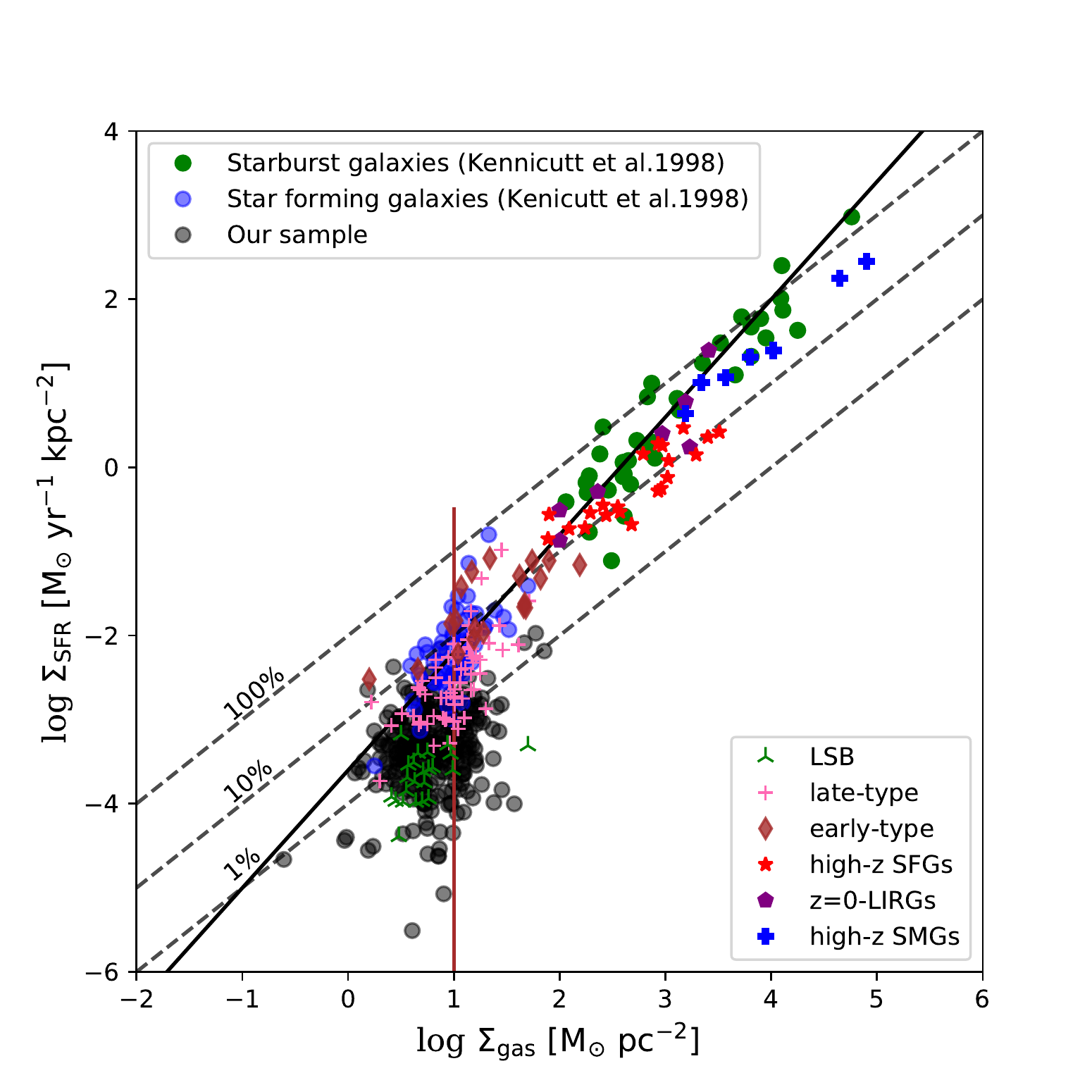}
\caption{The Kennicutt-Schmidt Law. 
Our LSBGs sample are black solid circles.
The blue dots and green dots are the star-forming galaxies and starburst galaxies from \citet{1998ApJ...498..541K}.
All the other symbols in the low-right box  are collected by \citet{2011ApJ...733...87S}.
The black solid line is the Kennicutt-Schmidt Law, three dotted lines showing the SFE of 100\%,10\%,1\% in a timescale of star formation of $10^{8}$ yr. 
The brown dashed line is the upper boundary of low gas surface density of 10 $\rm M_{\odot}pc^{-2}$.
}
\label{ks}
\end{figure*}

\begin{figure*}[!ht]
\centering
\includegraphics[scale=0.7]{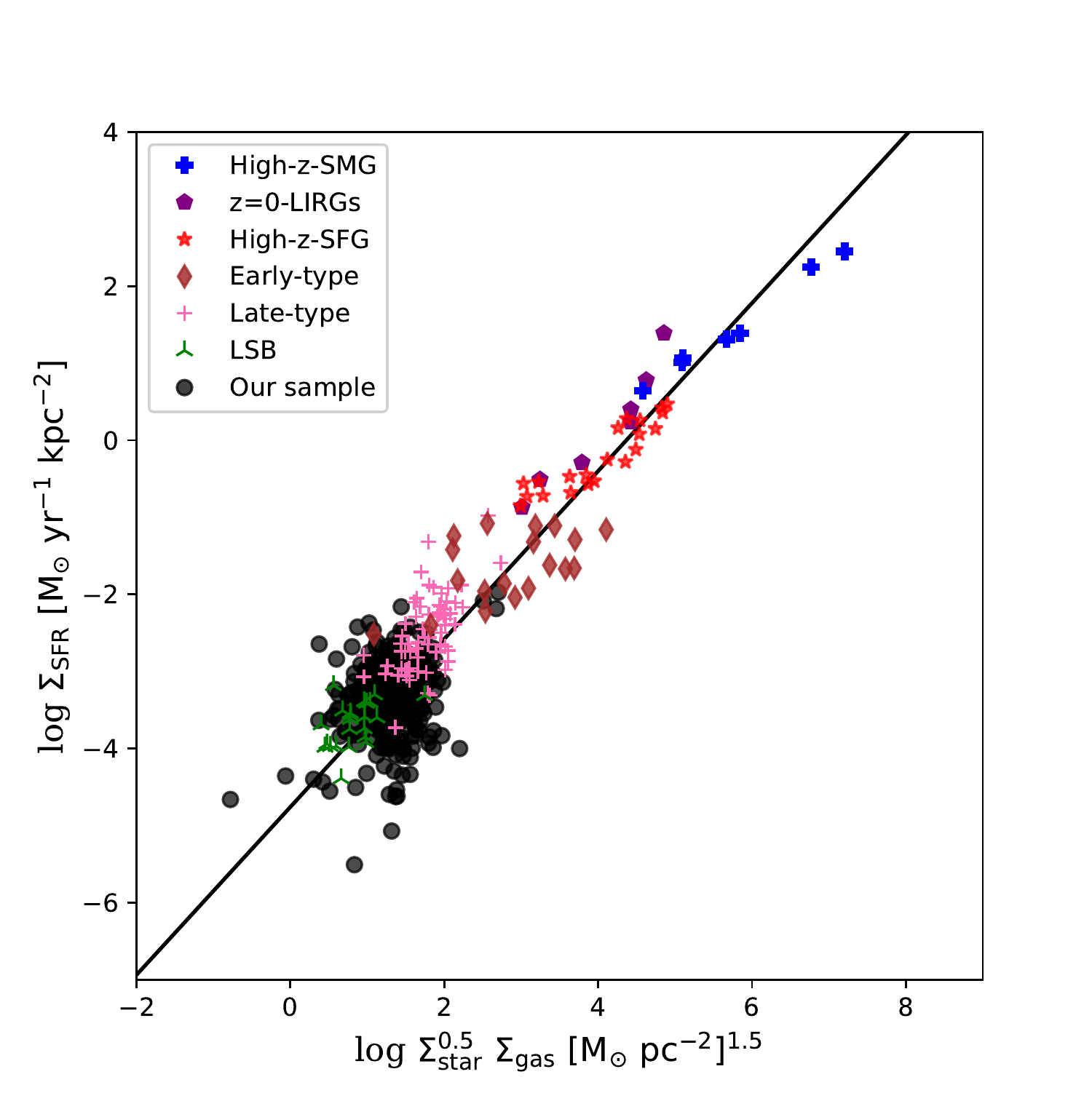}
\caption{The Extended Schmidt law. 
All the symbols are the same as those in Figure \ref{ks}.
 The black solid dots are LSBGs in our sample.
The black line is the best-fit line from high-z-SMG to LSB,  which is named as the extended Schmidt law \citep{2018ApJ...853..149S}.
}
\label{eks}
\end{figure*}

\section{The $\ha$ Flux Catalog}
Table \ref{tab:tabletwo} presents the $\ha$ fluxes, SFRs, H\Rmnum{1} masses, and stellar masses of 357 spring LSBGs.
83$\%$ of them show positive $\ha$ emission detection.
We also append the new results of 111 fall LSBGs from \citet{2018ApJS..235...18L} with a new continuum subtraction method to Table \ref{tab:tabletwo}.
The columns of Table \ref{tab:tabletwo} are as the following:

Column 1. The entry number of the Arecibo General Catalog (AGC).

Column 2. The semi-major axis of elliptical photometry (a). The elliptical isophotes are employed to fit the galaxy images by the IRAF ellipse package.
The semi-major axis  adopts $r_{25}$, where the surface brightness magnitude reaches to 25 $\rm mag\ arcsec^{-2}$ in units of kpc.

Column 3. The ellipticity of galaxy is defined as 1-(b/a).  a and b are semi-major axis and mini-major axis, respectively.

Column 4. The logarithm of the $\ha$ flux and errors.
The $\ha$ flux are total flux enclosed within the ellipse with a semi-major axis of $r_{25}$ after a series of flux corrections. The unit is $\rm erg\ s^{-1} cm^{-2}$.

Column 5. The logarithm of the SFR ($\rm M_{\sun} yr^{-1}$).
The SFR  is calculated from $\rm SFR_{H\alpha}(M_{\sun}\ yr^{-1}) = 7.9 \times 10^{-42}[L(H\alpha)](erg\ s^{-1})$,
where L(H$\alpha$) is the extinction corrected H$\alpha$ luminosity \citep{1998ApJ...498..541K}.

Column 6. The logarithm of the SFR surface density ($\rm M_{\sun} yr^{-1} kpc^{-2}$).
The elliptical photometry area is used to calculate the star formation surface density ($\rm \Sigma_{SFR}=SFR/\pi ab$).

Column 7. The logarithm of the H\Rmnum{1} mass taken from the $\rm \alpha$.40 catalog.
H\Rmnum{1} mass is computed via the standard formula of $\rm M_{HI}=2.356 \times 10^{5} D_{Mpc}^{2} S_{21}$,
where $\rm D_{Mpc}$ is the distance, and the $\rm S_{21}$ is the integrated H\Rmnum{1} line flux density of the source in $\rm Jy\ km\ s^{-1}$\citep{2011AJ....142..170H}.

Column 8. The logarithm of the H\Rmnum{1} gas surface density ($\rm M_{\sun} pc^{-2}$).
As $\rm r_{HI}/r_{25}$ is mostly constant (1.7$\pm$0.5) and shows weak dependence on the types from S0 to Im, 
we adopt 1.7 times optical radii of $r_{25}$ as the H\Rmnum{1} radii \citep{1997A&A...324..877B,2002A&A...390..863S,2015ApJ...808...66J}. The H\Rmnum{1} gas surface density is calculated as $\rm \Sigma_{HI}=M_{HI}/2.89\pi ab$.

Column 9. The stellar mass is derived from the r-band luminosity and g-r color using \citet{2003ApJS..149..289B}'s formula.

Column 10. The logarithm of the mass surface density ($\rm M_{\sun} pc^{-2}$). $r_{25}$ is used to calculate the mass surface density.

In order to check the reliability of our results, we compare the $\ha$ fluxes of LSBGs which also have been observed in the $\ha 3$ Survey \citep{2012A&A...545A..16G,2015A&A...576A..16G}.
In Figure \ref{comflux}, the blue solid circles are galaxies matched with the Virgo cluster \citep{2012A&A...545A..16G} and the red solid circles are galaxies matched with the Coma cluster \citep{2015A&A...576A..16G}.
We find most of the $\ha$ fluxes show a good agreement with $1\sigma$ uncertainty, and the mean value and the standard deviation of the differences between them is 0.11 and 0.34, respectively.

\section{Result and Discussion}
\noindent{\emph{SFR and H\Rmnum{1} Gas Distribution}}

Figure \ref{segsfr} shows the distributions of the SFR, the star formation efficiency of H\Rmnum{1} gas ($\rm SFE_{HI}=SFR/M_{HI}$), the surface density of the SFR and the mass of H\Rmnum{1} gas of different galaxies.
In each case, the black solid lines show the corresponding distributions for our LSBGs sample.
In panel (a) and (b), the solid blue lines and red lines show the distributions of the star-forming galaxies and starburst galaxies from \citet{1996AJ....112.1903Y} and \citet{2015ApJ...808...66J}, respectively.
In panel (c) and (d), the solid green lines and purple lines also represent the star-forming galaxies and starburst galaxies from \citet{1998ARA&A..36..189K}.

Comparing to the star-forming and starburst galaxies, LSBGs show a similar distribution of $\rm \Sigma_{HI}$, but both SFRs and $\rm SFE_{HI}$s of the LSBGs are lower than those of the star-forming galaxies by more than one order of magnitude, and far lower than those of the starburst galaxies. 
Furthermore, the $\rm \Sigma_{SFR}$ of LSBGs are about two order of magnitudes lower than that of the star-forming galaxies.
All these distributions indicate that the H\Rmnum{1} gas-rich galaxies do not mean higher SFRs and $\rm \Sigma_{SFR}$.\\
\hspace{1cm}\\
\emph{H\Rmnum{1} Mass, Stellar Mass and SFR }

Figure \ref{sfrmass} presents the ratio of the H\Rmnum{1} mass to the stellar mass versus the H\Rmnum{1} mass (a1, a2), SFR versus the stellar mass (b1, b2) and the ratio of the SFR to  H\Rmnum{1} mass versus the H\Rmnum{1} mass (c1, c2).

The left panels show the galaxies from the Virgo and Coma clusters\citep{2012A&A...545A..16G,2013A&A...553A..91F}, the dwarf galaxies from \citet{2012AJ....143..133H} and the galaxies from xGASS \citep{2017ApJS..233...22S,2018MNRAS.476..875C}, as a comparison.
The galaxies of the $\ha 3$ survey covering the region of the Virgo and Coma clusters are shown in brown and black.
Their SFRs are calculated from $\ha$ imaging.
The stellar mass of the $\ha3$ survey is derived from the i-band magnitudes and g-i color using the formula given by \citet{2003ApJS..149..289B}.
The purple pluses present the dwarf galaxies provided by  \citet{2012AJ....143..133H}  selected by crossmatching the $\alpha$.40 catalog, SDSS and Galaxy Evolution Explorer (GALEX).
The stellar mass and the SFRs of 80 dwarf galaxies are derived by fitting their UV-optical SEDs.
The H\Rmnum{1} masses of the galaxies in Coma cluster, the Virgo cluster and the dwarf galaxies are from ALFALFA too.
In Figure \ref{sfrmass}, we also show the galaxies in xGASS \citep{2018MNRAS.476..875C}, which is a sample  selected homogeneously by the stellar mass.
The xGASS is an extended program from $GALEX$ Arecibo SDSS Survey (GASS) \citep{2010MNRAS.403..683C,2011MNRAS.415...32S}.
The lower limit of the stellar mass extends from $\rm 10^{10}M_{\odot}$ to $\rm 10^{9} M_{\odot}$.
Figure \ref{sfrmass} only shows xGASS galaxies with H\Rmnum{1} detection.
The H\Rmnum{1} mass is from the Arecibo observation.
The stellar mass is from the SDSS DR7 Max Planck for Astrophysics/Johns Hopkins University catalog.
The SFR is calculated as described in \citet{2017MNRAS.466.4795J}.

The right panels show three types of our LSBGs in open circles and the comparison galaxies mentioned above in smaller pluses.
Our LSBGs are separated into three types according to their absolute B magnitudes: the giant ($\rm M_{B} < -19\ Mag$), the intermediate ($\rm -19 \leqslant M_{B} \leqslant -17\ Mag$) and the dwarf LSBGs ($\rm M_{B} > -17\ Mag$)\citep{2019MNRAS.483.1754D}.

In panel (a1), the solid line represents the H\Rmnum{1} mass equals the stellar mass.
The ratio of the H\Rmnum{1} mass to the stellar mass decreases as the stellar mass increase. 
This is consistent with the result in \citet{2010MNRAS.403..683C,2011MNRAS.415.1797C,2011MNRAS.411..993F,2012ApJ...756..113H}.
In panel (a2), our LSBGs follow the similar relation of the comparison galaxies in panel (a1).
Panel (b) shows the SFR vs. the stellar mass. 
The relation between the SFR and the stellar mass is crucial for understanding the star formation history and evolution of galaxies.
The SFR increases with the stellar mass, which is the so-called star-forming main sequence \citep{2004MNRAS.351.1151B,2007ApJ...660L..43N,2009MNRAS.400..154B,2011A&A...533A.119E}. 
In the panel (b1) and (b2), the two solid lines are the same and are the fitting line of the comparison galaxies in panel (b1).
The stellar mass has good relation with the SFRs in all types of galaxies.
The LSBGs also follow this relation.
In panels (c1, c2), two solid lines are the median value of  $\rm SFR/M_{HI}$ of the comparison galaxies in panel (c1).
Comparing panel (c1) to (c2), most of the LSBGs present relatively lower values of $\rm SFR/M_{HI}$ than other ALFALFA-selected galaxies in panel(c1). From the definition of $\rm SFE_{HI}=SFR/M_{HI}$, we conclude that most of the LSBGs have lower star formation efficiency of H\Rmnum{1} gas.

\hspace{1cm}\\
\emph{Kennicutt-Schmidt (K-S) Law  }

Until now it is still hard to detect CO emission line in LSBGs.
Only a few of LSBGs detected molecular gas \citep{2001ApJ...549L.191M, 2003ApJ...588..230O, 2005AJ....129.1849M, 2010A&A...523A..63D,2017AJ....154..116C}.
The relation between $\rm \Sigma_{SFR}$ and the H\Rmnum{1} gas surface density ($\rm \Sigma_{HI}$) is shown in Figure \ref{hiks}.
The black solid circles are LSBGs in our sample.
The orange stars represent the LSBGs from \citet{2009ApJ...696.1834W}.
The blue open circles ($\rm \Sigma_{HI}$) are the star-forming galaxies from \citet{1998ApJ...498..541K}.
As in Figure \ref{segsfr}, the $\rm \Sigma_{HI}$ of both the LSBGs and the star-forming galaxies distribute in the similar range. 
The $\rm \Sigma_{SFR}$ of the star-forming galaxies are higher than those of LSBGs. Both kinds of galaxies are distinguished by the different $\rm \Sigma_{SFR}$.
A positive relation between $\rm \Sigma_{HI}$ and $\rm \Sigma_{SFR}$ for star-forming galaxies (blue open circles) seems to exist, but none for LSBGs (black solid circles).

Though we have no molecular mass of LSBGs, $\rm \Sigma_{H_{2}}$ can be roughly estimated  according to $\rm \Sigma_{SFR}$ \citep{2008AJ....136.2846B}.
Based on such an estimation, the $\rm \Sigma_{H_{2}+HI}$ of our LSBGs is very close to $\rm \Sigma_{HI}$, which is consistent with the previous assumption that H\Rmnum{1} dominates the gas content in our LSBGs \citep{2018ApJS..235...18L}.
Therefore, it is reasonable to employ the $\rm \Sigma_{HI}$ instead of the $\rm \Sigma_{gas}$ for LSBGs in the plot of the K-S law.

Figure \ref{ks} shows the relation between the SFR surface density ($\rm \Sigma_{SFR}$) and the gas surface density ($\rm \Sigma_{gas}$).
The blue and green solid circles represent the star-forming galaxies and starburst galaxies from \citet{1998ApJ...498..541K}, respectively.
Six types of galaxies are collected by \citet{2011ApJ...733...87S}. 
They are late-type galaxies (pink pluses), early-type galaxies (brown thin diamonds), LSB galaxies (green tri-down) from \citet{2009ApJ...696.1834W}, local luminous infrared galaxies (z=0-LIRGs, purple hexagons), high-redshift star-forming galaxies (High-z SFGs, red stars) and high-redshift merging submillimeter galaxies (High-z SMG, blue filled pluses).
The black solid circles are LSBGs in our sample.
The black solid line is the K-S law and the black dotted lines are SFE from 1$\%$ to 100$\%$,in a timescale of star formation of $10^{8}$ yr .

As shown in Figure \ref{ks}, combining the molecular gas, most galaxies (the local and high-redshift star-forming galaxies, starburst galaxies, luminous infrared galaxies, late-type and early-type galaxies, and submillimeter galaxies) follow the K-S law. 
However LSBGs obviously  deviate from the K-S law which is based on the star-forming and starburst galaxies. With the median value of $\rm SFE_{HI}$ around 1$\%$, the LSBGs are galaxies with low star formation efficiency. 
Also, most of the LSBGs have a gas surface density lower than the brown upper limit line of the low-density region \citep{2012ARA&A..50..531K}, which correspond to low-density systems.

The K-S law is actually an empirical relationship between the SFR surface density and the gas surface density based on a sample of 61 nearby spiral and 36 infrared-selected starburst galaxies \citep{1998ApJ...498..541K}.
\citet{2008AJ....136.2846B} have shown how the SFR, H\Rmnum{1} and $\rm H_{2}$ surface densities related to each other at sub-kpc resolution in 18 nearby galaxies.
Most galaxies show a good relation between $\rm\Sigma_{H_{2}}$ and $\rm\Sigma_{SFR}$, but they show little or no correlation between $\rm\Sigma_{HI}$ and $\rm\Sigma_{SFR}$.
Hence the galaxies dominated by $\rm H_{2}$ gas (i.e. star-forming galaxies, starburst galaxies, and LIRGs\ldots ) follow the K-S law. When galaxy (i.e. LSBGs and dwarf galaxies) have higher fraction of H\Rmnum{1} gas, they will deviate the K-S law.

Our LSBGs are selected from ALFALFA H\Rmnum{1} survey. 
From \citet{2012ApJ...756..113H}, the galaxies detected by the ALFALFA survey bias to the gas-rich system. 
Compared to the optically selected galaxies, the H\Rmnum{1}-selected population has overall higher SFR and sSFRs at a given stellar mass, but lower $\rm SFE_{HI}$. Similar to the parent sample, our LSBGs are also tend to lower SFE. Furthermore, from Figure \ref{sfrmass} (c2), even comparing to other H\Rmnum{1}-selected galaxies from ALFALFA survey, the LSBGs show much lower $\rm SFE_{HI}$. 
The lower  the $\rm SFE_{HI}$ the farther away the LSBGs deviate from the  K-S law.

We use $\rm \Sigma_{HI}$ instead of $\rm \Sigma_{gas}$ in Figure \ref{ks} .
If taking the molecular gas into consideration, the LSBGs should shift to the right, and even further from the K-S law.
As shown in Figure \ref{segsfr}, comparing to the star-forming galaxies, LSBGs show a similar distribution in $\rm \Sigma_{HI}$, but $\rm \Sigma_{SFR}$ of the LSBGs are much lower than those of the star-forming galaxies.
The LSBGs deviate from the K-S law because their $\rm \Sigma_{SFR}$ is low.
The right panel of Figure \ref{photo} shows the star-forming regions of five example LSBGs.
The distribution of the star-forming region is widely sparse, so the filling factor of star formation region is lower in LSBGs \citep{2009ApJ...696.1834W}.
This would lead to an  lower $\rm \Sigma_{SFR}$ when averaging over the entire galaxy.
We will further study the filling factor of LSBGs in the future work.
\\
\hspace{1cm}\\
\emph{Extended Schmidt Law }

The K-S law does not hold for the entire range of gas densities, especially in the low gas density.
In order to solve such a problem, \citet{2011ApJ...733...87S,2018ApJ...853..149S} added the stellar mass surface density into the K-S law, after evaluating the importance of existing stars in the whole galaxy's history.
Figure \ref{eks} shows the extended Schmidt law as a black line from \citet{2018ApJ...853..149S}, which is an empirical relation between the $\Sigma^{0.5}_{star}\Sigma_{gas}$ and $\Sigma_{SFR}$ from different types of galaxies.
All symbols are same as those in Figure \ref{ks}.
It contains our LSBGs (black solid circles),  late-type galaxies (pink pluses), early-type galaxies (brown thin-diamond), LSB galaxies (green tri-down) from \citet{2009ApJ...696.1834W}, local luminous infrared galaxies (z=0-LIRGs, purple hexagon), high-redshift star-forming galaxies (High-z SFGs, red star) and high-redshift merging sub-millimeter galaxies (High-z SMG, blue filled pluses).
The different apertures are adopted for the different type galaxies to derive the SFR and $\rm \Sigma_{SFR}$.
\citet{2011ApJ...733...87S} pointed out that the different apertures do not affect the extended Schmidt law.
Because the SFR, gas, and stellar mass are measured within same aperture, the galaxies would move along the extended Schmidt law without large offsets when different apertures are applied.

Although our LSBGs deviated obviously from the K-S law, they follow the extended Schmidt law.
The quantitative analysis shows that the LSBGs present a median offset of 0.041 dex from the extended Schmidt law, which is much smaller than that of 0.844 dex from the K-S law.
This confirms that the extended Schmidt law is more suitable for the extremely low gas density environment, such as our gas-rich LSBGs. 

Star formation can be described as a conversion of gas to a star over a timescale.
The K-S law suggests that the gas mass surface density is the only factor in regulating the SFR surface density.
However, the extended Schmidt law shows that the star formation is well correlated with both the stellar mass and the gas ($\Sigma_{SFR} \propto \Sigma_{star}\Sigma_{gas}$). Our observation data of LSBGs confirms the superiority of the extended Schmidt law \citep{2011ApJ...733...87S,2018ApJ...853..149S}.

Compared to $\Sigma_{star}$, $\Sigma_{gas}$ show little correlation with SFE in H\Rmnum{1}-dominated region \citep{2008AJ....136.2782L}.
$\rm \Sigma_{star}$ is much better than $\Sigma_{gas}$ in predicting the SFR of the H\Rmnum{1}-dominated galaxies, which is in agreement with the result in \citet{1998ApJ...493..595H}.
This is also supported by Figure \ref{sfrmass}, in which the stellar mass correlates well with the SFR for both the normal star-forming galaxies and LSBGs (b1,b2), but the LSBGs deviate obviously from the normal galaxies in the relation of the H\Rmnum{1} mass and the SFR (c1,c2).  
 
In fact, the star formation is a complex process, so many  works \citep{2008AJ....136.2782L,2011ApJ...733...87S,2017A&A...608A..24R,2018ApJ...853..149S} suggest that the star formation can be regulated by stellar mass through its gravity and gas pressure.
Recent work shows the crucial importance of feedback from earlier generation of stars in setting up the  pressure in the interstellar medium and affecting future star formation.
Star formation is regulated in such a manner can, in principle, be the reason behind the extended Schmidt law.

Though the LSBGs follow the extended Schmidt law, they still present a relatively large scatter, which requires us to take more factors into consideration in the future studies.

\section{Summary}

We perform a narrow $\ha$-band imaging survey of LSBGs selected from the spring region of the 40$\%$ ALFALFA extragalactic H\Rmnum{1} survey.
Our sample contains 357 spring LSBGs, and is observed with the Xinglong 2.16 m telescope, which belongs to National Astronomical Observatories, Chinese Academy of Sciences (NAOC).
We update the process of data reduction, especially continuum subtraction.
We present a catalog of the $\ha$ fluxes and derived parameters of 357 spring LSBGs and 111 fall LSBGs in \citet{2018ApJS..235...18L} with new continuum subtraction.

Compared to the star-forming galaxies, LSBGs have similar H\Rmnum{1} surface densities but have much lower SFRs and SFR surface densities.
The relation between the $\rm \Sigma_{SFR}$ and $\rm \Sigma_{gas}$ shows that our H\Rmnum{1}-dominated LSBGs obviously deviate from the K-S law of the star-forming galaxies and starburst galaxies, possibly because of their low-density environment, low star formation efficiency, and low filling factor of star-forming regions.
After taking the stellar mass into consideration, the LSBGs follow the extended Schmidt law well,
with a mean offset of 0.041 dex, compared to a mean offset of 0.844 dex from the K-S law.
Our results suggest that the extended Schmidt law can suit for the star formation in the low-density environment.

\begin{acknowledgements}
We thank the referee for constructive comments and suggestions.
This project is supported by the National Key R$\&$D Program of China (No. 2017YFA0402704), and the National Natural Science Foundation of China (grant No. 11733006, 11403037, 11225316, 11173030, 11303038,  11403061, and U1531245).
We acknowledge the support of the staff of the Xinglong 2.16 m telescope. 
This work was partially supported by the  Open  Project  Program  of  the  Key  Laboratory  of  Optical  Astronomy,  National Astronomical Observatories, Chinese Academy of Sciences.

We acknowledge the work of the entire ALFALFA collaboration team in observing, flagging and extracting the catalog of galaxies used in this work. The ALFALFA team at Cornell is supported by NSF grant AST-0607007 and AST-1107390 and by grants from the Brinson Foundation.
We thank the useful SDSS database and the MPA/JHU catalogs.
Funding for the SDSS  has been provided by the Alfred P. Sloan Foundation, the Participating Institutions,
the National Science Foundation, the U.S. Department of Energy, the National Aeronautics and Space Administration, the Japanese Monbukagakusho, the Max Planck Society, and the Higher Education Funding Council for England.
\end{acknowledgements}

%
\startlongtable
\begin{deluxetable*}{rcrrrrrrrr}
\tablecaption{The Observed Sample of LSBGs \label{tab:tableone} }

\tablehead { 
\colhead{AGC} & \colhead{$\rm \mu_{0}(B)$} & \colhead{g}& \colhead{r}&\colhead{R.A.} & \colhead{Dec.} &   \colhead{z} & \colhead{Dist} & \colhead{Filter} &  \colhead{Date} \\
\colhead{}& \colhead{$\rm mag\ arcsec^{-2}$} & \colhead{mag}   & \colhead{mag} &\colhead{J2000} & \colhead{J2000}& \colhead{}&\colhead{Mpc}&\colhead{} & \colhead{} \\
\colhead{(1)} &\colhead{(2)} &\colhead{(3)} &\colhead{(4)} &\colhead{(5)} &\colhead{(6)} &\colhead{(7)} &\colhead{(8)} &\colhead{(9)} &\colhead{(10)} 
           }
\startdata
           4542    &     22.75    &     14.97    &     14.52    &       08:42:53    &      +25:04:11    &    0.0173    &      75.5    &       Ha3    &  20130410   \\
           4626    &     23.68    &     15.79    &     15.37    &       08:51:01    &      +24:19:09    &    0.0091    &      41.0    &       Ha2    &  20131230   \\
           4797    &     23.48    &     14.13    &     13.69    &       09:08:11    &      +05:55:39    &    0.0044    &      20.9    &       Ha2    &  20150421   \\
           5633    &     22.68    &     13.86    &     13.44    &       10:24:40    &      +14:45:23    &    0.0046    &      22.6    &       Ha2    &  20131230   \\
           5716    &     22.93    &     15.25    &     15.07    &       10:31:43    &      +25:18:26    &    0.0043    &      21.1    &       Ha2    &  20160205   \\
           5758    &     23.00    &     16.41    &     16.12    &       10:36:13    &      +13:26:57    &    0.0099    &      45.0    &       Ha2    &  20131230   \\
           6122    &     22.95    &     15.54    &     15.15    &       11:03:32    &      +11:07:04    &    0.0213    &      96.3    &       Ha4    &  20140405   \\
           6248    &     23.68    &     15.72    &     15.28    &       11:12:52    &      +10:12:00    &    0.0043    &      17.5    &       Ha2    &  20140406   \\
           6287    &     23.39    &     16.60    &     16.29    &       11:16:06    &      +23:54:37    &    0.0209    &      93.9    &       Ha4    &  20130412   \\
           6486    &     23.15    &     15.68    &     15.31    &       11:29:12    &      +11:51:55    &    0.0108    &      48.7    &       Ha2    &  20140403   \\
\enddata
\tablenotetext{}{(This table is available in its entirety in a machine-readable form in the online journal.)}
\end{deluxetable*}
%

\startlongtable
\begin{deluxetable*}{rrrccccccc}
\tablecaption{The Star Formation Properties of LSBGs \label{tab:tabletwo}}
\tablehead{ 
\colhead{AGC}&  \colhead{$\rm r_{25}$} & \colhead{ellipse} & \colhead{$\rm log F(H\alpha)$}  & \colhead{log$\rm (SFR)$} & \colhead{$\rm log \Sigma_{sfr}$} & \colhead{$\rm log M_{HI}$} & \colhead{$\rm log \Sigma_{HI}$} & \colhead{$\rm log M_{*}$} & \colhead{$\rm log \Sigma_{star}$} \\
\colhead{}&  \colhead{Kpc}&\colhead{}& \colhead{$\rm erg\ cm^{-2}\ s^{-1}$}& \colhead{$\rm M_{\odot}yr^{-1}$} & \colhead{$\rm M_{\odot}yr^{-1}Kpc^{-2}$}& \colhead{$\rm M_{\odot}$}&  \colhead{$\rm M_{\odot}pc^{-2}$}& \colhead{$\rm M_{\odot}$} & \colhead{$\rm M_{\odot}pc^{-2}$}\\
 \colhead{(1)} &\colhead{(2)} &\colhead{(3)} &\colhead{(4)} &\colhead{(5)} &\colhead{(6)} &\colhead{(7)} &\colhead{(8)} &\colhead{(9)} &\colhead{(10)}
} 
\startdata
4542     &      14.25   &      0.18   &      $-13.34^{+0.10}_{-0.14}$     &      -0.60     &       -3.32     &       9.72    &       0.54      &       9.97    &       1.25    \\ 
4626     &      5.72    &      0.15   &      $-13.70^{+0.09}_{-0.12}$     &      -1.50     &       -3.44     &       9.39    &       0.99      &       9.06    &       1.12    \\ 
4797     &      4.37    &      0.1    &      $-13.57^{+0.25}_{-0.67}$     &      -1.96     &       -3.69     &       8.79    &       0.60      &       9.17    &       1.44    \\ 
5633     &      6.2     &      0.2    &      $-13.18^{+0.15}_{-0.22}$     &      -1.50     &       -3.49     &       9.35    &       0.90      &       9.32    &       1.33    \\ 
5716     &      3.67    &      0.41   &      $-13.05^{+0.04}_{-0.04}$     &      -1.42     &       -2.82     &       9.11    &       1.25      &       8.35    &       0.95    \\ 
5758     &      4.28    &      0.2    &      $-13.97^{+0.12}_{-0.16}$     &      -1.69     &       -3.35     &       8.78    &       0.66      &       8.7     &       1.04    \\ 
6122     &      15.17   &      0.32   &      $-13.87^{+0.14}_{-0.20}$     &      -0.93     &       -3.62     &       9.86    &       0.71      &       9.85    &       1.16    \\ 
6248     &      3.42    &      0.2    &      \nodata                   &      \nodata &       \nodata &                8.26    &     0.33   &       8.39    &       0.93    \\ 
6287     &      11.23   &      0.2    &      \nodata                   &      \nodata &       \nodata &                9.75    &     0.79   &       9.29    &       0.78    \\ 
6486     &      8.74    &      0.22   &      \nodata                   &      \nodata &       \nodata &                9.5     &     0.77   &       9.19    &       0.92    \\ 
\enddata
\tablenotetext{}{(This table is available in its entirety in a machine-readable form in the online journal.)}
\end{deluxetable*}

\bibliographystyle{aasjournal}
\bibliography{test}{}
\end{document}